\documentclass[prb,twocolumn,floatfix,nobibnotes,footinbib,superscriptaddress,
showpacs]{revtex4}
\usepackage[latin1]{inputenc}
\usepackage{amsmath}
\usepackage{graphicx}
\usepackage{psfrag}
\newcommand{\abs}[1]{\ensuremath{\lvert#1\rvert}}
\newcommand{\vc}[1]{\ensuremath{\mathbf{#1}}}

\begin{document}

\title{Dual Fermion Approach to Susceptibility of Correlated Lattice Fermions}

\author{S. Brener}
\affiliation{I. Institute of theoretical Physics, University of Hamburg, 20355
Hamburg, Germany}
\author{H. Hafermann}
\affiliation{I. Institute of theoretical Physics, University of Hamburg, 20355
Hamburg, Germany}
\author{A. N. Rubtsov}
\affiliation{Department of Physics, Moscow State University, 119992 Moscow,
Russia}
\author{M. I. Katsnelson}
\affiliation{Institute for Molecules and Materials, Radboud
University of Nijmegen, 6525 AJ Nijmegen, The Netherlands}
\author{A. I. Lichtenstein}
\affiliation{I. Institute of theoretical Physics, University of Hamburg, 20355
Hamburg, Germany}
\affiliation{Kavli Institute for Theoretical Physics, University of California,
Santa Barbara, California 93106, USA
}

\date{\today}

\begin{abstract}
In this paper, we show how the two-particle Green function (2PGF) can be obtained within the framework of the Dual Fermion approach. This facilitates the calculation of the susceptibility in strongly correlated systems where long-ranged non-local correlations cannot be neglected. We formulate the Bethe-Salpeter equations for the full vertex in the particle-particle and particle-hole channels and introduce an approximation for practical calculations. The scheme is applied to the two-dimensional Hubbard model at half
filling. The spin-spin susceptibility is found to strongly  increase for the wavevector $\vc{q}=(\pi,\pi)$, indicating the antiferromagnetic instability. We find a suppression of the critical temperature compared to the mean-field result due to the incorporation of the non-local spin-fluctuations.
\end{abstract}


\maketitle

\section{Introduction}

Strongly correlated electron systems exhibit some of the most intriguing
features known to condensed matter physics, including high-temperature
superconductivity, heavy-fermion behavior, different kinds of electronic phase
transitions, etc. \cite{mott,anderson1,anderson2,dwave,hewson}. In most of the 
cases, a conventional band theory \cite{AM,VK} is not sufficient to describe
these properties. Essentially a many-body treatment is necessary, correlation
effects being too strong to be described by the standard perturbation theory.
Due to the complexity of this problem the physics of strongly correlated systems
has proven to be theoretically challenging. One of the successful routes to the
description of strongly correlated systems is the Dynamical Mean Field Theory
(DMFT) \cite{DMFT1, DMFT2}. It is commonly accepted now that this approach
typically catches the most essential correlation effects, e.g., the physics of
the Mott-Hubbard transition \cite{DMFT1, DMFT2}. The method was implemented
successfully into realistic electronic structure calculations
\cite{Anisimov97,LK98}, which is now a standard tool in the microscopic theory
of strongly correlated systems \cite{Kotliar2006}. However, there are many
phenomena for which \emph{non-local} correlations are important and often the
relevant correlations are even long-ranged. Among them are the Luttinger-liquid
formation in low-dimensional systems \cite{anderson2,Mahan}, non-Fermi-liquid
behavior due to van-Hove singularities  in two dimensions
\cite{vanHove1,vanHove2}, the physics near quantum critical points
\cite{Sachdev}, or d-wave pairing in high-$T_c$ superconductors \cite{dwave}.

The requirement to incorporate spatial correlations of strongly
correlated fermions into present theories has triggered several
efforts to go beyond DMFT. In the Dynamical Vertex Approximation
\cite{Tos07} and similar approaches \cite{Kus06,Sle06}, a
diagrammatic expansion around the DMFT solution is performed. In
Ref. \onlinecite{Tos07}, the authors introduce a scheme based on
the assumption of the locality of the fully irreducible vertex of
the lattice problem. In the framework of their approximation this
vertex is extracted from the Anderson impurity model. The
reducible vertex can be obtained from the parquet equations which
is subsequently used to calculate the non-local self-energy. The
Green function is updated and the parquet equations are solved
again until self-consistency is reached. In the work cited above a
simplified version of this algorithm was implemented, where
instead of the parquet equations the Bethe-Salpeter equation (BSE) in a
particular channel was solved without performing a self-consistent
calculation. Hence, this approach does not go beyond the usual
DMFT approximation for calculating the two-particle Green
functions (2PGFs). Practically, to go well beyond DMFT within this
approach, the parquet equations, which form a complicated system
of coupled integral equations need to be solved repeatedly, which
requires a sizeable numerical effort.

A principally new scheme with a fully renormalized expansion
called Dual Fermion Approach has been proposed recently
\cite{Rub08}. It is based on the introduction of new variables in
the path integral representation. This approach yields very
satisfactory results already for the lowest-order corrections,
while the schemes proposed in Refs. \cite{Tos07, Kus06, Sle06}
operate with infinite diagrammatic series and require the solution
of complicated integral equations. A scheme similar to the Dual
Fermion approach has been discussed earlier in terms of Hubbard
operators \cite{Kot04}, but has not been used for practical calculations.

In many cases, it is desirable to calculate the 2PGF which
provides insight into the inter-particle interactions and
two-particle excitations of the system and corresponding
instabilities. In the framework of DMFT the 2PGFs are usually
calculated under the assumption that the irreducible vertex part
in a given channel is local and it is taken to be equal to the
corresponding irreducible vertex of the impurity problem. This
assumption is empirical and does not hold in the general case
\cite{comment}.

In this paper we describe how to perform calculations of the 2PGF
within the Dual Fermion approach. We will show that already a
simple ladder approximation gives very reasonable results for the
Hubbard model at half filling, for example a reduction of the critical temperature of the antiferromagnetic instability compared to the DMFT result is obtained already in the framework of the single-site calculations. It occurs at the N\'{e}el temperature which
is in agreement with quantum Monte Carlo results \cite{QMCchi} for the same
parameters. It is remarkable that to obtain these results it is
sufficient to take the bare dual fermion vertex as the irreducible
vertex in the Bethe-Salpeter equations and only the lowest-order
diagram for the dual self-energy. Such effectiveness is achieved
by transforming the original interacting problem to so-called dual
fermion variables. By this we are able to include the local
contribution to the self-energy into a bare propagator of the dual
fermions and achieve a much faster convergence of the perturbation
series. The outcome of the scheme is the 2PGF in the original
variables restored from the dual 2PGF with the help of an exact
relation \cite{Haf07}.

This paper is organized as follows: In Section II we review the
dual fermion formalism in the general multiband formulation. Section III
gives a general overview of the exact equations for the 2PGFs in
different channels. We then discuss various possible
approximations which should be used for different purposes. We
further derive the exact relation between 2PGFs in the dual and
original variables. Section IV is devoted to the application of
the approach to the two-dimensional (2D) Hubbard model at half
filling and to the discussion of those results. In section V we
give the conclusions and summary.

\section{Dual Fermion Formalism}

The goal is to find an (approximate) solution to the general multiband
problem described by the imaginary time action
\begin{eqnarray}
S[c^\ast,c]\!&=&\!\!
-\!\!\!\!\!\sum_{\omega{\bf k}\sigma
mm^\prime}\!\!\! c^\ast_{\omega{\bf k}\sigma m}\left((i\omega+\mu){\bf 1}
-h_{{\bf k}\sigma}\right)_{mm^\prime}c_{\omega{\bf k}\sigma
m^\prime}\nonumber\\
&+& \sum_{i} H_{\text{int}}[c_i^\ast,c_i]\ .
\end{eqnarray}
Here $h_{{\bf k}\sigma}$ is the one-electron part of the
Hamiltonian, $\omega_n =(2n+1)\pi /\beta, n=0,\pm 1,...$ are the
Matsubara frequencies, $\beta $ and $\mu $ are the inverse
temperature and chemical potential, respectively, $\sigma
=\uparrow ,\downarrow $ labels the spin projection, $m, m'$ are
orbital indices and $c^{\ast },c$ are Grassmann variables. The
index $i$ labels the lattice sites and the ${\bf k}$-vectors are
quasimomenta.
For applications it is important to note that $H_{\text{int}}$ can be
\emph{any} type of interaction. The only requirement is that it is local within
the multiorbital atom or cluster. Consider, for example, the general Coulomb
interaction
\begin{equation}
H_{\text{int}}[c_i^\ast,c_i] = \frac{1}{4} \int\limits_0^\beta d\tau\, U_{1234} c^\ast_{i1} c^\ast_{i2} c_{i4} c_{i3}\ ,
\end{equation}
where $U$ is the general symmetrized Coulomb vertex and e.g.
$1\equiv\{\omega_1 m_1\sigma_1\}$ comprehends frequency-, orbital- and spin
degrees of freedom and summation over these states is implied.

The idea of the dual fermion approach is to reformulate the lattice
problem in terms of noninteracting impurities with their interaction replaced
by a coupling to auxiliary, so-called dual fermions.

As in the DMFT, we therefore introduce a local impurity problem in
the form
\begin{eqnarray}
S_{\text{imp}}[c^\ast,c]\!&=&\!\!
-\!\sum_{\omega\sigma mm^\prime}
c^\ast_{\omega\sigma
m}\left((i\omega+\mu){\bf 1}-\Delta_{\omega\sigma}\right)_{mm^\prime} c_{
\omega\sigma m^\prime}\nonumber\\
&+& H_{\text{int}}[c^\ast,c]\ ,
\end{eqnarray}
where $\Delta$ is an as yet unspecified hybridization matrix
describing the interaction of the impurity cluster with an
electronic bath.

Since we anticipate the decoupling of the interacting sites, we employ the
locality of $\Delta$ to formally rewrite the original lattice problem
in the following form:
\begin{eqnarray}
S[c^\ast,c]&=&\sum_i S_{\text{imp}}[c^\ast_{\omega
i\sigma m},c_{\omega i\sigma m}]\nonumber\\
&-& \sum_{\omega{\bf k}\sigma mm^\prime} c^\ast_{\omega{\bf k}\sigma
m} \left(\Delta_{\omega\sigma} -
h_{{\bf k}\sigma}\right)_{mm^\prime} c_{\omega{\bf k}\sigma
m^\prime}\ .\nonumber\\
\label{eq::action_rew}
\end{eqnarray}
We introduce spinors ${\bf c}_{\omega{\bf k}\sigma}=(\ldots
,c_{\omega{\bf k}\sigma m},\ldots)$, ${\bf c}^\ast_{\omega{\bf k}\sigma}=(\ldots
,c^\ast_{\omega{\bf k}\sigma m},\ldots)$. Omitting indices, the dual fermions
are introduced via a Gaussian identity which in matrix-vector notation reads
\begin{eqnarray}
&&\int \exp\left(-{\bf f}^*\hat{A}{\bf f} -
{\bf f}^*\hat{B}{\bf c}
- {\bf c}^*\hat{B}{\bf f}  \right) \mathcal{D}[{\bf f}^*,{\bf f}] =\nonumber\\
&&\det(\hat{A})\ \exp\left({\bf c}^*\hat{B}\hat{A}^{-1}\hat{B}{\bf c}\right)\ .
\end{eqnarray}
This identity is valid for arbitrary complex matrices $\hat{A}$ and
$\hat{B}$.
Choosing
\begin{eqnarray}
A&=&
g_{\omega\sigma}^{-1}\left(\Delta_{\omega\sigma}-h_{{\bf k}\sigma}\right)^{-1}g_{
\omega\sigma}^{-1}\ ,\nonumber\\
B&=&g_{\omega\sigma}^{-1}\ ,
\end{eqnarray}
where $g_{\omega\sigma}$ is the Green function matrix of the local impurity
problem in orbital space $(m,m^\prime)$, we obtain:
\begin{eqnarray}
&&S[{\bf c}^*,{\bf c},{\bf f}^*,{\bf f}] = \sum_i S_{\text{site},i} +\nonumber\\
&&\sum_{\omega{\bf k}\sigma}\left[{\bf f}^*_{\omega{\bf k}\sigma}\
g_{\omega\sigma}^{-1}\ \left(\Delta_{\omega\sigma} -
h_{{\bf k}\sigma}\right)^{-1}\!
g_{\omega\sigma}^{-1}\ {\bf f}_{\omega{\bf k}\sigma}\right]\ .
\end{eqnarray}
Hence the coupling between sites is transferred to a coupling to the auxiliary
fermions:
\begin{equation}
S_{\text{site},i}= S_{\text{imp}}[{\bf c}^*_i,{\bf c}_i] +
{\bf f}^*_{\omega i\sigma}\ g_{\omega\sigma}^{-1}{\bf c}_{\omega i\sigma} +
{\bf c}^*_{\omega i\sigma}\ g_{\omega\sigma}^{-1}{\bf f}_{\omega
i\sigma}\ .
\label{eq::Ssite}
\end{equation}
Note that since $g_{\omega\sigma}$ is local and the last term appears under a
sum over all states labeled by ${\bf k}$, the summation can be replaced by the
equivalent summation over all sites.

By transferring the inter-site coupling to interacting auxiliary fermions, only local degrees of freedom of the original fermions remain, so that we are able to integrate out the lattice fermions for each site separately:
\begin{eqnarray}
&&\int\exp\left(-S_{\text{site}}[{\bf c}_i^\ast,{\bf c}_i,{\bf f}_i^\ast,{\bf
f}_i]
\right)\mathcal{D}[{\bf c}_i^\ast,{\bf c}_i] =\nonumber \\
&& Z_{\text{imp}}e^{-\left(\sum_{\omega\sigma} {\bf f}^\ast_{\omega i\sigma}\
g^{-1}_{\omega\sigma}\ {\bf f}_{\omega i\sigma} +
V_i[{\bf f}_i^\ast,{\bf f}_i]\right)}\ .
\label{eq::def_V}
\end{eqnarray}
Formally this can be done up to all orders and in this sense the
transformation to the dual fermions is exact. The above equation
may be viewed as the defining equation for the dual potential
$V[{\bf f}^\ast,{\bf f}]$. Expanding both sides of Eqn.
(\ref{eq::def_V}) and equating the resulting expressions by order,
one finds that the dual potential in lowest order approximation is
given by
\begin{equation}
V[{\bf f}^\ast,{\bf f}] = \frac{1}{4} \sum\limits_i \gamma^{(4)}_{1234}
{\bf f}_{i1}^\ast
{\bf f}_{i2}^\ast {\bf f}_{i4} {\bf f}_{i3} + \ldots
\end{equation}
where
\begin{eqnarray}
\gamma^{(4)}_{1234} &=& g_{11^\prime}^{-1}g_{22^\prime}^{-1} \left[
\chi^{\text{imp}}_{1^\prime 2^\prime 3^\prime 4^\prime} -
\chi^{\text{imp}, 0}_{1^\prime 2^\prime 3^\prime 4^\prime} \right]g_{3^\prime
3}^{-1}g_{4^\prime 4}^{-1}\ , \nonumber \\
&&\chi^{\text{imp}, 0}_{1234} = g_{14}g_{23}
- g_{13} g_{24}
\label{eq::gamma4}
\end{eqnarray}
is the exact four-point reducible vertex for the original fermions which plays
the
role of the bare effective two-particle interaction for the dual fermions. The
local two-particle Green function of the impurity model is defined as
\begin{equation}
\chi^{\text{imp}}_{1234}=\frac{1}{Z_{\text{imp}}}\int c_{1}c_{2}
c_{3}^\ast c_{4}^\ast
\exp\left(-S_{\text{imp}}[{\bf c}^\ast,{\bf c}]\right)
\mathcal{D}[{\bf c}^\ast,{\bf c}]\ .
\label{eq::2pgf_imp}
\end{equation}
The dual action now depends on dual variables only and can be written as
\begin{equation}
S_{\text{d}}[{\bf f}^\ast,{\bf f}] = -\sum_{\omega{\bf k}\sigma}
{\bf f}^\ast_{\omega{\bf k}\sigma}
(G^{\text{d},0}_{\omega{\bf k}\sigma})^{-1}
{\bf f}_{\omega{\bf k}\sigma} + \sum_i V[{\bf f}_i^\ast,{\bf f}_i]\ .
\label{eq::dual_action}
\end{equation}
The bare dual Green function is given by
\begin{equation}
G^{\text{d},0}_{\omega{\bf k}\sigma} = -g_{\omega\sigma}\left[
g_{\omega\sigma}+\left(\Delta_{\omega\sigma} -
h_{{\bf k}\sigma}\right)^{-1}\right]^{-1} g_{\omega\sigma}\ .
\label{eq::Gdbare}
\end{equation}

As we shall see below, for a properly chosen $\Delta$, the DMFT result is
already recovered within the simplest (zero-order) approximation for the dual
potential $V$. In order to obtain the nonlocal corrections to the DMFT, we thus
need to calculate the dual self-energy up to higher orders in $V$.
This is achieved by performing a regular diagrammatic series expansion of the
dual action, Eqn. (\ref{eq::dual_action}).

\begin{figure}[t]
\begin{center}
\begin{tabular}{ccc}
 \includegraphics[scale=0.12,angle=0]{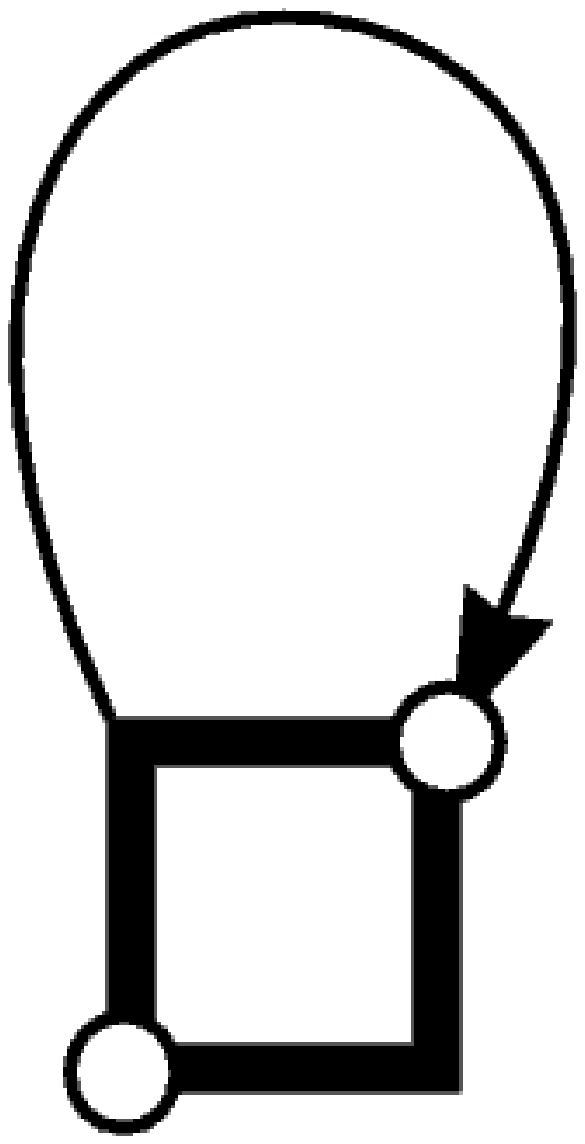} & \hspace{4em}
 \includegraphics[scale=0.12,angle=0]{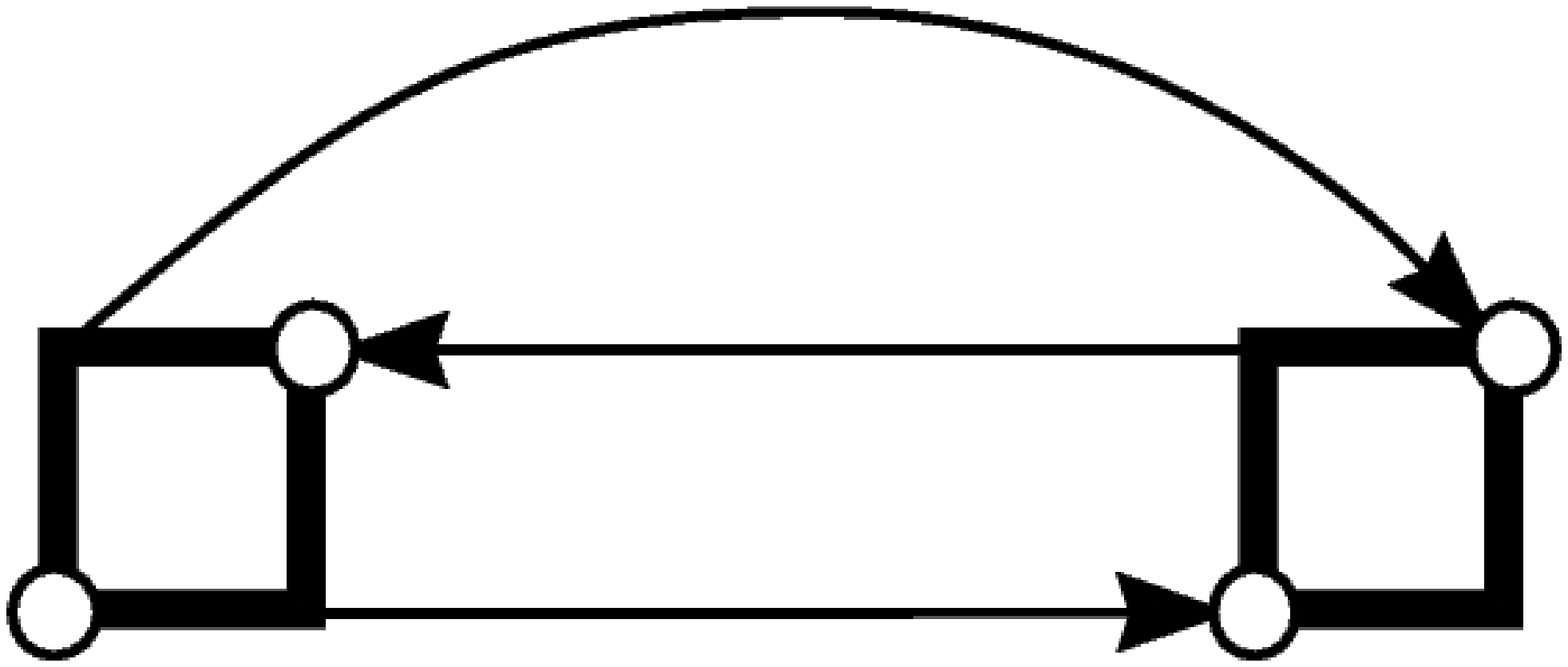}
\end{tabular}
\end{center}
\caption{The first two lowest order diagrams for the dual
self energy $\Sigma^d$.}\label{fig::diagrams}
\end{figure}

We note that the only approximation is that in practice the perturbation series expansion and the series for the dual potential need to be terminated at some point. Here we consider the first two lowest order skeleton diagrams for $\Sigma_{\text{d}}$, constructed from the irreducible vertices and the dual Green function as lines. The use of skeleton diagrams ensures that the resulting theory is conserving according to the
Baym-Kadanoff criterion \cite{BK, Rub08}. The diagrams considered here are shown
in Fig. \ref{fig::diagrams}. The lowest order diagram is local while the next
diagram already gives a nonlocal contribution to the self energy.

So far we have not established a condition for $\Delta$, which was so far an
arbitrary quantity. We require that the first diagram (Fig.
\ref{fig::diagrams}a) in the expansion of the dual self-energy should be equal
to zero at all frequencies. Since $\gamma^{(4)}$ is local, we can use the
condition $\sum_{{\bf k}}G_{{\bf k}\omega }^{d}=0$.
In the simplest approximation, which corresponds to non-interacting dual
fermions, the full dual Green function is replaced by the corresponding bare
Green function and the above condition can be reduced to
\begin{equation}
\sum_{{\bf k}} \left[g_{\omega\sigma}+\left(\Delta_{\omega\sigma} -
h_{{\bf k}\sigma}\right)^{-1} \right]^{-1}=0
\end{equation}
which is equivalent to the self-consistency condition for the
hybridization function in DMFT.

Thus, the case of non-interacting dual fermions corresponds to the full DMFT
result for the original fermions. In order to go beyond DMFT it is sufficient to
include a finite number of non-vanishing skeleton dual diagrams.
In this paper all calculations are performed using only the lowest order
non-vanishing skeleton diagram shown in Fig. \ref{fig::diagrams}b).
We postpone the explanation of the resulting numerical calculation procedure to
section IV.

The fact that we have employed an exact identity to transform to the dual
variables has the important consequence that we can establish an
exact relation between the lattice Green function and the dual
Green function. This is important to obtain the Green functions for the
original fermions without loss of information. We will further use this fact to
establish the relation between original and dual 2PGFs. To this end, the
partition function of the lattice is written in the two equivalent forms
\begin{eqnarray}
Z&=&\int\exp\left(-S[{\bf c}^\ast,{\bf c}]\right) \mathcal{D}[{\bf
c}^\ast,{\bf c}]
=\nonumber\\ && Z_f \int\int
\exp\left(-S[{\bf c}^\ast,{\bf c},{\bf f}^\ast,{\bf f}]\right)
\mathcal{D}[{\bf f}^\ast,{\bf f}]\mathcal{D}[{\bf c}^\ast,{\bf c}]\ ,\nonumber\\
\label{eq::partitionfunction}
\end{eqnarray}
where
\begin{equation}
Z_f = \prod\limits_{\omega{\bf k}\sigma}
\det\left[g_{\omega\sigma}\left(\Delta_{\omega\sigma}-h_{{\bf
k}\sigma}\right)g_{
\omega\sigma} \right]\ .
\end{equation}
In the following, we introduce the quantities
\begin{equation}
L_{\omega{\bf k}\sigma} = \left(\Delta_{\omega\sigma} -
h_{{\bf k}\sigma}\right)^{-1} g_{\omega\sigma}^{-1}
\label{eq::l}
\end{equation}
and
\begin{equation}
R_{\omega{\bf k}\sigma} = g_{\omega\sigma}^{-1} \left(\Delta_{\omega\sigma} -
h_{{\bf k}\sigma}\right)^{-1} 
\label{eq::r}
\end{equation}
which are matrices in orbital space.
By taking the functional derivative of the partition function, Eqn.
(\ref{eq::partitionfunction}), w.r.t. the Hamiltonian, the exact
relation between the dual and lattice Green functions can now conveniently be
written as
\begin{equation}
G_{\omega{\bf k}\sigma}=\left(\Delta_{\omega\sigma} - h_{{\bf
k}\sigma}\right)^{-1} + L_{\omega{\bf k}\sigma}\, 
 G^{\text{d}}_{\omega{\bf k}\sigma}\, R_{\omega{\bf k}\sigma}
\label{eq::Glat}
\end{equation}
where the Green functions are defined via the imaginary time
path integral as
\begin{eqnarray}
G_{12} &=& -\frac{1}{Z}\int c_{1}c_{2}^\ast
\exp\left(-S[{\bf c}^\ast,{\bf c}]\right)
\mathcal{D}[{\bf c}^\ast,{\bf c}] \nonumber \\
G^{\text{d}}_{12} &=& -\frac{Z_f}{Z}\int f_{1}f_{2}^\ast
\exp\left(-S[{\bf f}^\ast,{\bf f}]\right)
\mathcal{D}[{\bf f}^\ast,{\bf f}]\ . \nonumber \\
\label{eq::Gdual}
\end{eqnarray}

\section{Two-Particle Green Function.}

\subsection{Relation between 2PGF's in the original and dual variables.}

Analogously to the single-particle Green function, the connection between the original and dual 2PGF's can be
established by taking the second derivative with respect to $h_{\nu\mu}$ of the partition function $Z$ expressed in two different
ways: Eqn. (\ref{eq::partitionfunction}). Here and further the small Greek
letters denote the set $\{\omega,\mathbf k,\sigma,m\}$ and the summation over
repeated indices is implied. The 2PGFs are defined in a similar way as in
Eqn. (\ref{eq::2pgf_imp}):
\begin{equation}
\chi_{\lambda\mu\nu\rho}=\frac{1}{Z}\int c_{\lambda}c_{\mu}
c_{\nu}^\ast c_{\rho}^\ast
\exp\left(-S[{\bf c}^\ast,{\bf c}]\right)
\mathcal{D}[{\bf c}^\ast,{\bf c}]\ .
\label{eq::2pgf_original}
\end{equation}
\begin{equation}
\chi^{\text d}_{\lambda\mu\nu\rho}=\frac{Z_f}{Z}\int f_{\lambda}f_{\mu}
f_{\nu}^\ast f_{\rho}^\ast
\exp\left(-S_{\text d}[{\bf f}^\ast,{\bf f}]\right)
\mathcal{D}[{\bf f}^\ast,{\bf f}]\ .
\label{eq::2pgf_dual}
\end{equation}

Varying the first expression for $Z$ yields by definition:
\begin{equation}
\frac{1}{Z}\frac{\delta^2Z}{\delta h_{\rho\lambda}\delta h_{\nu\mu}} = \chi_{\lambda\mu\nu\rho}.
\end{equation}
The most illustrative way to vary the second expression for $Z$ from
Eqn. (\ref{eq::partitionfunction}) is to write
$$
\frac{1}{Z}\frac{\delta^2Z}{\delta h_{\rho\lambda}\delta h_{\nu\mu}}=\frac{1}{Z}\frac{\delta}{\delta h_{\rho\lambda}}
\left(\frac{\delta Z}{\delta h_{\nu\mu}}\right)=
-\frac{1}{Z}\frac{\delta}{\delta h_{\rho\lambda}}\left(Z\,G_{\mu\nu}\right).
$$
In the last expression we should use Eqn. (\ref{eq::Glat}) for $G$ understanding
it as being diagonal in frequency, momenta and spin indices.
From these two expressions we further obtain $\chi_{\lambda\mu\nu\rho} =
G_{\lambda\rho}G_{\mu\nu}-\frac{\delta G_{\mu\nu}}{\delta h_{\rho\lambda}}$. 
Using the identity
$$
\frac{dA^{-1}_{\alpha\beta}(x)}{dx}=-A^{-1}_{\alpha\gamma}\frac{dA_{\gamma\delta}}{dx}A^{-1}_{\delta\beta}
$$
for the derivative of an inverse matrix with respect to a parameter and the
Eqn. (\ref{eq::Glat}) we can rewrite the functional derivative of $G$ as
\begin{eqnarray}
\lefteqn{ \frac{\delta G_{\mu\nu}}{\delta h_{\rho\lambda}} =
\left(\Delta-h\right)^{-1}_{\lambda\nu}\left(\Delta-h\right)^{-1}_{\mu\rho}
+ L_{\mu\mu'}\frac{\delta G_{\mu'\nu'}^{d}}{\delta h_{\rho\lambda}} R_{\nu'\nu}
} \nonumber \\
&& \left(\Delta-h\right)^{-1}_{\lambda\nu}\left[L\, G_d\, R\right]_{\mu\rho}
+\left[L\, G_d\, R\right]_{\lambda\nu}
\left(\Delta-h\right)^{-1}_{\mu\rho}.
\label{eq::2pgfs_connection_inter2}
\end{eqnarray}


The derivative in the last term in Eqn. (\ref{eq::2pgfs_connection_inter2}) is
evaluated using the definition of the dual Green function, Eqn.
(\ref{eq::Gdual})
and the identity

\begin{equation}
\frac{\delta}{\delta h_{\rho\lambda}}\left(\frac{Z}{Z_f}\right) =-L_{
\lambda\lambda'}\, \frac{Z}{Z_f}G^d_{\lambda'\rho'}\, R_{\rho'\rho }\ .
\end{equation}


Thus,
\begin{equation}
\frac{\delta G_{\mu'\nu'}^{d}}{\delta h_{\rho\lambda}}=L_{\lambda\lambda'}\,
(G_{\lambda'\rho'}^{d}G^{d}_{\mu'\nu'}-\chi^d_{\lambda'\mu'\nu'\rho'})\,
R_{\rho'\rho}\ .
\label{eq::2pgfs_connection_inter3}
\end{equation}



Using Eqn. (\ref{eq::2pgfs_connection_inter3}) in
(\ref{eq::2pgfs_connection_inter2}), we obtain the final result:
\begin{eqnarray}
\chi_{\lambda\mu\nu\rho}
&=&\left[\left(\Delta-h\right)^{-1}\otimes\left(\Delta-h\right)^{-1}\right]_{
\lambda\mu\nu\rho} \nonumber \\
&&+\left[\left(\Delta-h\right)^{-1}\otimes\left[L\,
G_d\, R\right]\right]_{\lambda\mu\nu\rho} \nonumber\\
&&+\left[\left[L\, G_d\, R\right]
\otimes\left(\Delta-h\right)^{-1}\right]_{\lambda\mu\nu\rho}\nonumber \\
&&+ L_{\lambda\lambda'}\, L_{\mu\mu'}\,
\chi^d_{\lambda'\mu'\nu'\rho'}\, R_{\nu'\nu}\, R_{\rho'\rho}\ ,
\label{eq::2pgfs_connection}
\end{eqnarray}

where $(A\otimes B)_{\lambda\mu\nu\rho}\equiv
A_{\lambda\rho}B_{\mu\nu}-A_{\lambda\nu}B_{\mu\rho}$ is the antisymmetrized
direct matrix product.

The formula (\ref{eq::2pgfs_connection}) can be (symbolically) rewritten as
\begin{equation}
\chi-G\otimes G = L\, L\, (\chi^d-G^d\otimes G^d)\, R\, R,
\label{eq::realchi}
\end{equation}


which shows quite clearly that the two particle excitations in the original and
dual variables are identical.

\subsection{Bethe-Salpeter Equations for Dual Fermions}

As is clearly seen from the end of the previous section, the
problem of calculating the 2PGF in the original variables reduces
to the problem of calculating it for the dual fermions. In this
section we review the standard Bethe-Salpeter equation approach
to this problem and justify the approximations we use in the
calculations, the details of the latter will be given in Section
IV.

In what follows we denote the 2PGF as $\chi$ and the vertices as
$\Gamma$. Different sub- and superscripts will be used to
distinguish between different channels and types of vertices.
Small Greek letters stand for the set ${\omega, k, m}$ (or with
appropriate changes $\omega\to\tau$ and/or $k\to x$ if the
calculations are done in imaginary time or real space), spin will
be taken into account explicitly. We also omit the index ``d'' for
``dual'' unless it is necessary for the sake of clarity. Otherwise
all quantities used are dual by default.

The full vertex $\Gamma$ is defined by:
\begin{eqnarray}
\chi_{\lambda\mu\nu\rho}^{\sigma\sigma'}&=&G_{\lambda\lambda'}^{\sigma}G_{
\mu\mu' }^{\sigma'}\Gamma_{\lambda'\mu'\nu'\rho'}^{\sigma\sigma'}
G_{\nu'\nu}^{\sigma'}G_{\rho'\rho}^{\sigma} \nonumber \\
&&+G_{\lambda\rho}^{\sigma}G_{\mu\nu}^{\sigma'}-G_{\lambda\nu}^{\sigma}G_{
\mu\rho } ^ {\sigma'}\delta_{\sigma\sigma'}\ .
\end{eqnarray}
For this vertex one can write three different Bethe-Salpeter
equations:
\begin{eqnarray}
\Gamma_{\lambda\mu\nu\rho}^{\sigma\sigma'}&=&\Gamma_{\lambda\mu\nu\rho}^{pp,
\sigma\sigma'\!\text{,ir}}
+\xi\Gamma_{\lambda\mu\mu'\lambda'}^{pp,\sigma\sigma'\!\text{,ir}}G_{
\lambda'\rho'}^{\sigma}G_{\mu'\nu'}^{\sigma'}\Gamma_{\rho'\nu'\nu\rho}^{
\sigma\sigma'}, \nonumber\\
\\
\label{eq::bse_pp}
\Gamma_{\lambda\mu\nu\rho}^{\sigma\bar{\sigma}}&=&\Gamma_{\lambda\mu\nu\rho}^{
ph1 , \sigma\bar{\sigma}\text{,ir}}
-\Gamma_{\lambda\nu'\nu\lambda'}^{ph1,\sigma,\bar{\sigma}\text{,ir}}G_{
\lambda'\rho' }
^{\sigma}G_{\mu'\nu'}^{\bar{\sigma}}\Gamma_{\rho'\mu\mu'\rho}^{\sigma\bar{
\sigma } } , \nonumber\\
\\
\label{eq::bse_ph1}
\Gamma_{\lambda\mu\nu\rho}^{\sigma\sigma'}&=&\Gamma_{\lambda\mu\nu\rho}^{ph0,
\sigma\sigma'\!\text{,ir}}
-\Gamma_{\lambda\rho'\lambda'\rho}^{ph0,\sigma\sigma''\!\text{,ir}}G_{
\lambda'\nu'}^{\sigma''}G_{\mu'\rho'}^{\sigma''}
\Gamma_{\nu'\mu\nu\mu'}^{\sigma''\sigma'}, \nonumber\\
\label{eq::bse_ph0}
\end{eqnarray}
where $\bar{\sigma}$ denotes $-\sigma$ and
$\xi:=1-\frac12\delta_{\sigma\sigma'}$.
These equations are written in different channels. The first one in the
particle-particle channel and the latter two in the particle-hole channel. The
two ph-channels differ from each other by the total spin (0,1) of the scattered
particle-hole pair. $\Gamma^{\text{ir}}$ stands for the irreducible vertex in
the corresponding channel.

Obviously, the problem of calculating $\Gamma$ is numerically
solvable if $\Gamma^{\text{ir}}$ is known. The simplest possible
approximation for any irreducible vertex is the bare four-point
interaction of dual fermions i.e. $\gamma^{(4)}$. In this paper we
will not go beyond this approximation for practical calculations.
Substituting it into Eqns. (\ref{eq::bse_pp},\ref{eq::bse_ph0})
we obtain approximations for the exact vertex $\Gamma$ in the
respective channel:

\begin{eqnarray}
\Gamma_{\lambda\mu\nu\rho}^{pp,\sigma\sigma'}&=&\gamma_{\lambda\mu\nu\rho}^{
(4)\sigma\sigma'}
+\xi\gamma_{\lambda\mu\mu'\lambda'}^{(4)\sigma\sigma'}G_{\lambda'\rho'}^{
\sigma}G_{\mu'\nu'}^{\sigma'}\Gamma_{\rho'\nu'\nu\rho}^{pp,\sigma\sigma'},
\nonumber\\
\\
\label{eq::bse_pp_apr}
\Gamma_{\lambda\mu\nu\rho}^{ph1,\sigma\bar{\sigma}}&=&\gamma_{\lambda\mu\nu\rho}
^{ (4)\sigma\bar{\sigma}}
-\gamma_{\lambda\nu'\nu\lambda'}^{(4)\sigma\bar{\sigma}}G_{\lambda'\rho'}^{
\sigma }
G_{\mu'\nu'}^{\bar{\sigma}}\Gamma_{\rho'\mu\mu'\rho}^{ph1,\sigma\bar{\sigma}},
\nonumber\\
\\
\label{eq::bse_ph1_apr}
\Gamma_{\lambda\mu\nu\rho}^{ph0,\sigma\sigma'}&=&\gamma_{\lambda\mu\nu\rho}^{
(4)\sigma\sigma'}
-\gamma_{\lambda\rho'\lambda'\rho}^{(4)\sigma\sigma''}G_{\lambda'\nu'}^{\sigma''
}G_{\mu'\rho'}^{\sigma''}
\Gamma_{\nu'\mu\nu\mu'}^{ph0,\sigma''\sigma'}.\nonumber \\
\label{eq::bse_ph0_apr}
\end{eqnarray}

We expect this approximation to capture the essential features of
the problem if the channel is chosen properly. Indeed, as can be
seen from the previous works on dual fermions, the results for the
Green function turn out to be qualitatively correct already in the
lowest order in $\gamma$, this means that $\gamma$ is a good small
parameter in the problem. Our expectations are justified by
numerical calculations for the 2D Hubbard model at half-filling in the
next section. It is worth mentioning here that we do not expect
this simple approximation to give quantitatively correct results for
superconductivity, as the spin-spin correlations are known to be
of high importance for the d-wave superconductivity of the Hubbard
model. In order to approach this problem first the vertex in the
ph-channel should be calculated and the result should be used as
an approximation for the irreducible vertex in the pp-channel.
This work is in progress now.

We also want to comment on the standard DMFT approach for
calculating the 2PGFs. As has been mentioned before, the DMFT
result for Green function is reproduced from the dual fermion
approach if the bare dual Green functions are used. On the other
hand, if such an attempt is made for the 2PGF, i.e. the bare dual
Green functions and the bare dual four-point vertex $\gamma$ are
inserted in the Eqn. (\ref{eq::2pgfs_connection}), we \emph{do
not} reproduce the DMFT result. Instead what we obtain is:
\begin{equation}
\chi-G^{\text{D}}\otimes G^\text{D} = L\,L\, (G^
{\text{d},0}G^{\text{d},0}\gamma^{(4)}G^{\text{d},0}G^{\text{d},0})\, R\, R\ ,
\end{equation}
where $G^{\text{D}}=(g^{-1}+ (\Delta-h))^{-1}$ is the DMFT lattice
Green function. Using the relation
$(\Delta-h)^{-1}g^{-1}G^{\text{d},0}=-G^{\text{D}}$ (which can be
proven by straightforward calculation) leads to:
\begin{equation}
\chi-G^{\text{D}}\otimes
G^{\text{D}}=G^{\text{D}}G^{\text{D}}\gamma^{(4)}G^{\text{D}}G^{
\text{D}}
\end{equation}
In the last expression $\gamma^{(4)}$ - being the full vertex of the
Anderson impurity model - can be symbolically written in form of a
Bethe-Salpeter equation:
\begin{equation}
\gamma^{(4)}=\gamma^{\text{ir}}+\gamma^{\text{ir}}\,g\,g\,\gamma^{(4)}.
\end{equation}
If here we replace the local Green functions $g$ in the ladder by $G^{\text D}$,
we immediately restore the conventional result
for the 2PGF in DMFT approach: $\chi=\chi_0+\chi_0\gamma^{\text{ir}}\chi$ with $\chi_0=G^{\text D}\otimes G^{\text D}$. 
But this substitution can only be done ``by hand''
and as is obvious from the structure of the dual diagrams cannot
be reproduced automatically at any order of the perturbation
theory. This is an indication that the conventional way to
calculate the 2PGF is not $\Phi$-derivable and does not constitute
a systematic way to calculate the $k$-dependent vertex function.
This drawback is excluded in our approach. However, the DMFT
approach still works reasonably. One can see that by noticing that
$g=G^{\text{D}}-G^{\text{d},0}$ (derived straightforwardly). The
local part of $G^{\text{d},0}$ vanishes by construction and the
non-local part is generally small. Thus one can say that the DMFT
approach to calculating the 2PGF corresponds to considering the
first non-trivial diagram for the dual 2PGF with a small change of
a part of the internal Green functions of the vertex.

Finally we would like to mention that apart from $\gamma^{(4)}$ also higher
cumulants are present in the dual fermion
approach. We do not expect them to be important for calculating the 2PGF and do not take them into account in this paper.
However, the 6-point vertex $\gamma^{(6)}$
which plays the role of the effective 3-particle interaction for the dual
variables can turn out to be of crucial importance
for non-linear optics calculations.

\section{Application to the 2D Hubbard model}

Let us now turn to the Hubbard model. The 2D Hubbard model is described by the Hamiltonian 
\begin{equation}
H = -\sum\limits_{ij\sigma} t_{ij} c_{i\sigma}^\dagger c_{j\sigma} + U
\sum\limits_i
n_{i\uparrow}n_{i\downarrow}\ ,
\end{equation}
with the bare dispersion relation given by $h_\vc{k}=-2t(\cos k_x + \cos k_y)$. In the following the energy scale is fixed by setting the nearest neighbor hopping $t_{ij}=t=1$.
In this study, we restrict ourselves to the case of half-filling and postpone the case of finite doping including an analysis of the superconducting instability to a later publication.

It is known that at low temperatures strong antiferromagnetic correlations
develop in the 2D Hubbard model at half-filling. This is due to the perfect
nesting of the Fermi surface. Rigorously speaking, a transition to a
long-range ordered state at finite temperatures cannot occur since this would break the continuous spin 
symmetry, which is prohibited by the Mermin-Wagner theorem\cite{merminwagner}. For calculations however, one should keep in mind that a transition to the antiferromagnetic state is possible: Being bound to approximations, the fluctuations responsible for destroying the long-range order are not fully accounted for and the implications of the Mermin-Wagner theorem are not applicable. In this section we study how the antiferromagnetic instability emerges within our approach and how the transition temperature compares with that obtained within the DMFT.

The calculations are performed for the 2D Hubbard model at half-filling within the paramagnetic phase. In this publication, we restrict ourselves to the application of the single-site dual fermion approach. As a prerequisite for a better understanding on how the corrections to the DMFT emerge, we first briefly review the calculation procedure. Then, in order to illustrate the method and to underline some of the implications from the susceptibility calculations, we first present some results for single-particle properties obtained from DMFT and dual fermion (DF) calculations.
Afterwards, explicit expressions for the solution of the Bethe-Salpeter equation and the calculation of susceptibilities are introduced. Then we discuss the two-particle properties obtained using the ladder approximations to the 2PGF discussed in the preceding section. In order to compare the DF and DMFT results, we transform the dual susceptibility to the corresponding result for the original lattice fermions. The DF results are then contrasted to the corresponding DMFT results for different values of the on-site repulsion and different temperatures. We find considerable corrections to the DMFT from our DF calculations, in particular for large values of $U$.

\subsection{Dual Fermion calculations}

\begin{figure}[t]
\begin{center}
\includegraphics[scale=0.9,angle=0]{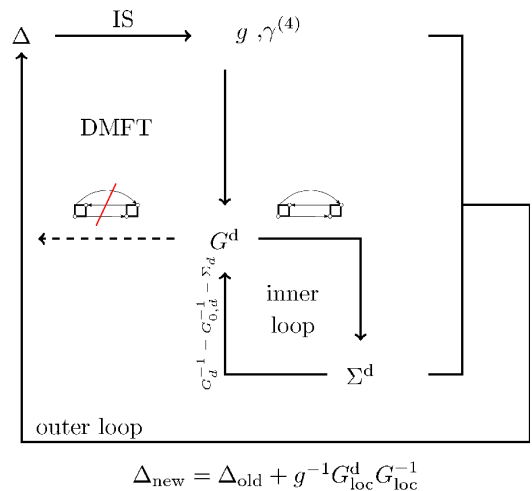}
\end{center}
\caption{(Color online) Illustration of the dual fermion calculation procedure.
}\label{fig::procedure}
\end{figure}

Since the method is rather new, let us first briefly review the calculation procedure. Additional information can be found in Refs. \onlinecite{Rub08} and \onlinecite{Haf07}.
Each calculation is started with a regular self-consistent DMFT calculation. This is achieved by requiring the local part of the bare dual fermion propagator (i.e. no corrections to the dual self-energy are taken into account as indicated by the crossed diagram in Fig. \ref{fig::procedure}) to be zero. We then calculate the vertex $\gamma^{(4)}$. This enables us to calculate an approximation to the dual self-energy by summing up the first diagrams in the perturbation series expansion. From this and the
bare dual Green function an approximation to the dual Green function $G^{\text{d}}$ is obtained via Dyson's equation, which is subsequently used in the diagrams. This inner loop is executed until self-consistency. This effectively replaces the diagrams depicted in Fig.\ref{fig::diagrams} by the corresponding skeleton diagrams. In general, the first-order diagram is now non-zero, in contrast to the DMFT result. From the condition that this diagram should be zero, we construct a new hybridization function, which serves as input for the calculation of a new local Green function and renormalized
vertex $\gamma^{(4)}$ in the impurity solver step. We repeat this outer loop until self-consistency is reached. The computational cost for the calculations aside from DMFT is comparable to the DMFT itself, whereby the calculation of the vertex is the computationally most expensive part. We use the continuous-time quantum Monte Carlo
impurity solver\cite{Rub05-1,Rub05-2,Rub05-3} for the solution of the impurity
problem and for the calculation of the vertex.

\subsection{Single-particle properties}

\begin{figure}[b]
\psfrag{x}{$\omega$}
\psfrag{y}{\hspace{-0.5em}DOS}
\begin{center}
\includegraphics[scale=0.32,angle=0]{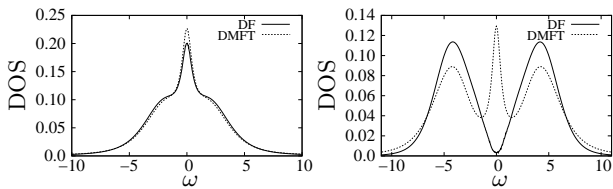}
\vspace{-2em}
\end{center}
\caption{Local density of states for the 2D Hubbard model at half-filling obtained within DMFT (dashed lines) and dual fermion calculations (solid lines) for $U/t=4$ (left) and $U/t=8$ (right) at inverse temperature $\beta=4.5$.
}\label{fig::dos}
\end{figure}

In this section we present some characteristic single-particle properties to illustrate our approach. All calculations have been performed for $U=4$ and $8$. Note that the bandwidth is $W=8t=8$.

To begin with, we present results for the local density of states (see Fig. \ref{fig::dos}). For $U=4$, we find qualitatively the same behavior within the DMFT and DF calculations, with the difference in the DF result being a suppression of the spectral weight around the Fermi energy compared to DMFT. For this value of $U$ the gap does not open, but in should be noted that in order to fully open a gap e.g. within the dynamical cluster approximation (DCA) requires large clusters of the order of 64 sites\cite{moukouri01,hushcroft01}. At $U=8$ however, we find a drastic change in the local density of states when the dual corrections are taken into account. We observe qualitatively similar changes in a broad temperature range above the critical temperature. In the DF calculations, the spectral weight around the Fermi level is strongly suppressed. This results in a pseudogap which persists up to higher temperatures. It is believed to be caused by non-local spin fluctuations, which are not present in the DMFT.

\begin{figure}[t]
\begin{center}
\includegraphics[scale=0.65,angle=0]{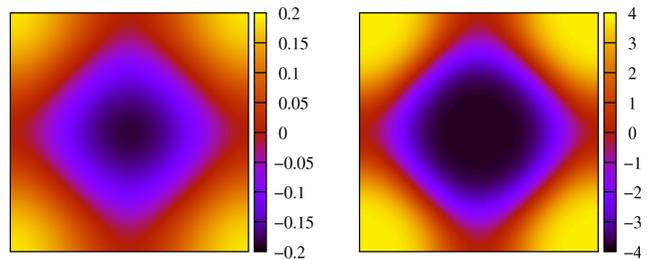}
\vspace{-2em}
\end{center}
\caption{(Color online) Contour plot of the real part of the lattice self-energy on the first Matsubara frequency $\Re \Sigma(\omega=\pi T,\vc{k})$ (centered at the $\Gamma$-point) for $U/t=4$ (left) and $U/t=8$ (right), calculated on a grid of $256\times 256$ k-points at inverse temperature $\beta=3.5$.
}\label{fig::resigma}
\end{figure}

\begin{figure}[t]
\begin{center}
\includegraphics[scale=0.65,angle=0]{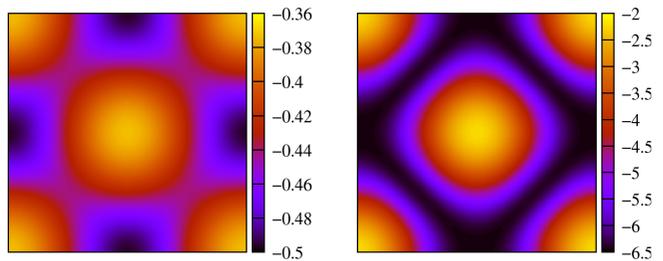}
\vspace{-2em}
\end{center}
\caption{(Color online) Contour plot of the imaginary part of the lattice self-energy $\Im \Sigma(\omega=\pi T,\vc{k})$ (centered at the $\Gamma$-point) for $U/t=4$ (left) and $U/t=8$ (right) at inverse temperature $\beta=3.5$.
}\label{fig::imsigma}
\end{figure}

This further suggests that non-local corrections to the DMFT self-energy become important in this parameter regime. To underline this picture, we show the real and imaginary parts of the k-dependent lattice self-energy on the first Matsubara frequency. Within DMFT, this quantity is just a constant, $\Sigma(\pi T,\vc{k})=\Sigma(\pi T)$. For the case of large on-site repulsion, we find a strong renormalization of the bare dispersion law $h_{\vc{k}}$ as shown in Fig. \ref{fig::resigma}. Also the imaginary part of the self-energy exhibits a strong k-dependence (see Fig \ref{fig::imsigma}). Qualitatively similar features are found for higher temperatures.

\begin{figure}[b]
\psfrag{x}{$\omega$}
\psfrag{y}{\hspace{-3.6em} {\scriptsize Spectral Function}}
\begin{center}
\includegraphics[width=0.25\textwidth,angle=0]{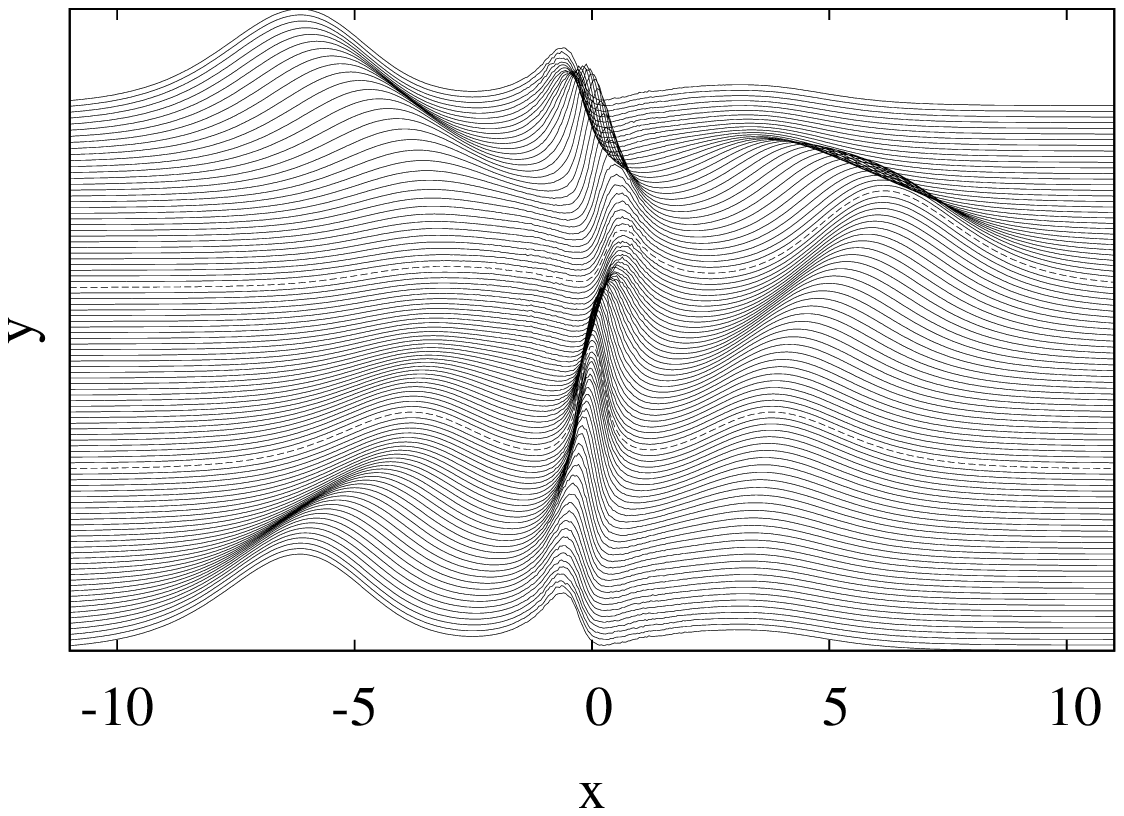}\includegraphics[width=0.25\textwidth,angle=0]{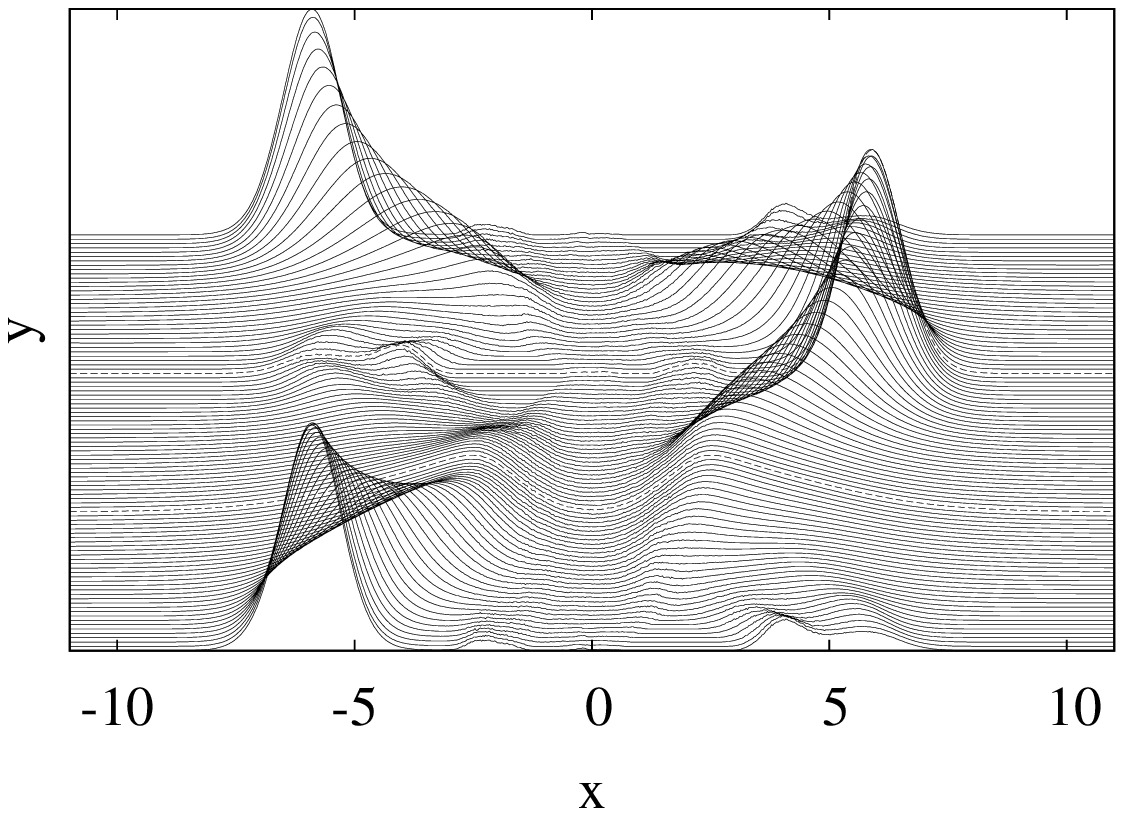}
\end{center}
\caption{Spectral function for the 2D Hubbard model at half-filling obtained within DMFT (left) and DF (right) calculations for $U/t=8$ at inverse temperature $\beta=4.5$. From bottom to top, the curves are plotted along the high-symmetry lines $\Gamma\rightarrow X\rightarrow M\rightarrow \Gamma$. The high symmetry points $X$ and $M$ are marked by dashed lines.
}\label{fig::sf}
\end{figure}

A quantity which is directly accessible to experimental observation via angular resolved photo emission spectroscopy is the single-particle spectral function $A(\vc{k},\omega)$. In Fig. \ref{fig::sf} we show $A(\vc{k},\omega)$ for $U=8$ and $\beta=4.5$ obtained from the DMFT and DF calculations, respectively. As expected from the local density of states, the DMFT and DF spectral functions are quite different. The most striking feature is the absence of a coherent peak at the Fermi level in the DF calculations. Our DF spectral function very well resembles the characteristic features of the one given in Ref. \onlinecite{stanescu04}. In this work, an approach based on Hubbard operators including spin-fluctuation corrections to the self-energy has been used. The self-energy was obtained from the self-consistent solution of a two-site dynamical cluster problem and is thus computationally more involved than our single-site approach. In addition, it treats the local (i.e. nearest-neighbor) correlations explicitly. This option is also present within the cluster-formulation of our approach.

\begin{figure}[bt]
\psfrag{x}{$i\omega$}
\psfrag{y}{$g_0(i\omega)$}
\begin{center}
\includegraphics[scale=0.34,angle=0]{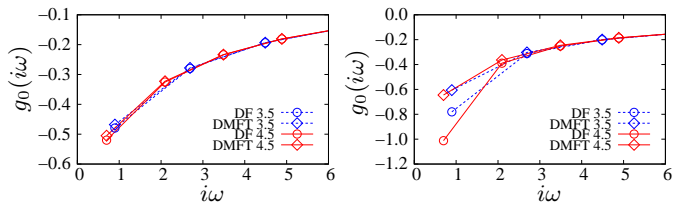}
\vspace{-2em}
\end{center}
\caption{(Color online) Comparison of the DMFT Weiss field $g_0(i\omega)=[i\omega+\mu-\Delta (i\omega)]^{-1}$ and its modification after a fully self-consistent dual fermion calculation (DF), for $U/t=4$ (left) and $U/t=8$ (right). The dashed curves correspond to the inverse temperature $\beta=3.5$ and the solid curves to $\beta=4.5$.
}\label{fig::wfield}
\end{figure}

Since obviously the DMFT insufficiently describes the physics leading to the appearance of the pseudogap, one might expect that the dynamical mean field, which describes the influence of the surrounding sites in a mean field manner, does not accurately reflect the surrounding in the real system. In such a case, the renormalization of the hybridization function $\Delta(i\omega)$ (cf. Fig. \ref{fig::procedure}) should become important. This can be seen in Fig. \ref{fig::wfield}, where we compare the Weiss field $g_0(i\omega)=[i\omega+\mu-\Delta(i\omega)]^{-1}$ from the DMFT and DF calculations. For $U=4$, the change is small and almost does not depend on temperature. On the other hand, for $U=8$, we find a strong renormalization, which significantly increases when the temperature is lowered. This reflects the fact that the local singlet formation prevails at low temperatures. Also note that the DMFT result hardly depends on temperature, in contrast to the DF results.

\subsection{Solution of the Bethe-Salpeter equation}

\begin{figure}[t]
\psfrag{s1}{$\sigma$}
\psfrag{s2}{$\bar{\sigma}$}
\psfrag{s3}{$\sigma'$}
\psfrag{s4}{$\sigma$}
\psfrag{s5}{$\sigma'$}
\psfrag{s6}{$\sigma''$}
\psfrag{g}{$\gamma$}
\psfrag{g0}[l]{$\Gamma^{\text{\tiny ph0}}$}
\psfrag{g1}[l]{$\Gamma^{\text{\tiny ph1}}$}
\psfrag{g2}[l]{$\Gamma^{\text{\tiny pp}}$}
\psfrag{x1}{$\vc{q}$}
\psfrag{x2}{ }
\psfrag{x3}{ }
\psfrag{x4}{ }
\psfrag{x5}{$\vc{q}$}
\begin{center}
\includegraphics[scale=0.375,angle=0]{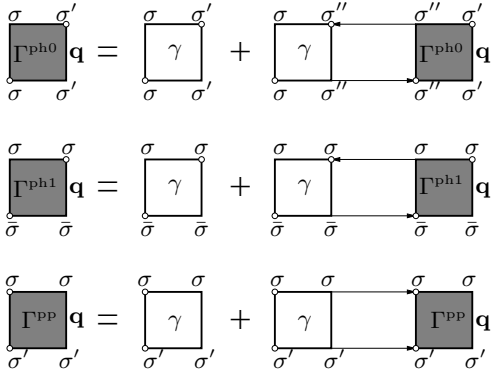}
\end{center}
\caption{The Bethe Salpeter equation in the three different
channels. In the calculations we considered the equations
for the particle-hole channel.}\label{fig::bse}
\end{figure}

In order to calculate the susceptibilities, we need to calculate the reducible
two-particle vertex by solving the Bethe-Salpeter equation.
Explicitly, the Bethe-Salpeter equations in the singlet and triplet
particle-hole channels (see Fig. \ref{fig::bse}) are
\begin{eqnarray}
\Gamma^{ph0,\sigma\sigma'}_{\omega\omega' \Omega} (\vc{q}) =
\gamma^{\sigma\sigma'}_{\omega\omega' \Omega}&-&
\frac{1}{\beta N^d}\sum\limits_{\omega''\sigma''}\sum\limits_{\vc{k}}\gamma^{
\sigma\sigma''}_{ \omega\omega''
\Omega}G^{\text{d}\sigma''}_{\omega''}(\vc{k})\times\nonumber \\
&&\times G^{\text{d}\sigma''}_{\omega''+\Omega}(\vc{k}+\vc{q})\, 
\Gamma^{ph0,\sigma''\sigma'}_{\omega''\omega'
\Omega}(\vc{q})\nonumber \\
\label{eq::bse1}
\end{eqnarray}
\begin{eqnarray}
\Gamma^{ph1,\sigma\bar{\sigma}}_{\omega\omega' \Omega} (\vc{q}) =
\gamma^{\sigma\bar{\sigma}}_{\omega\omega' \Omega}&-&
\frac{1}{\beta N^d}\sum\limits_{\omega''}\sum\limits_{\vc{k}}\gamma^{
\sigma\bar{\sigma}}_{\omega\omega'' \Omega} G^{\text{d}\bar{\sigma}}_{\omega''}(\vc{k})
\times\nonumber \\
&&\times G^{\text{d}\sigma}_{\omega'' +
\Omega}(\vc{k}+\vc{q})\, \Gamma^{ph1,\sigma\bar{\sigma}}_{
\omega''\omega^\prime\Omega}(\vc{q})\nonumber\ .\\
\label{eq::bse2}
\end{eqnarray}
Here $\gamma$ is the irreducible vertex and $G^{\text{d}\sigma}_{\omega}(\vc{k})$ denotes the fully self-consistent nonlocal dual fermion propagator. To be complete, we wrote the spin indices on the propagators. The calculations however are carried out for the paramagnetic case, i.e. $G^{\text{d}\sigma}=G^{\text{d}\bar{\sigma}}$. Capital letters $\Omega$ denote the bosonic and small letters fermionic Matsubara frequencies. $N$ is the linear dimension of the lattice and $d$ the dimension. Due to the ladder approximation, the resulting fully reducible vertex $\Gamma$ only depends on a single momentum $\vc{q}$.

The Bethe-Salpeter equation is solved iteratively. A starting guess for
the fully reducible vertex $\Gamma=\Gamma^{(0)}$ is inserted into the equation
and a new approximation $\Gamma^{(1)}$ is obtained. This process is repeated
until self-consistency. As a convergence criterion we require the difference
between two successive iterations to be smaller than some predetermined accuracy
$\epsilon$, i.e $\Vert \Gamma^{(n+1)}-\Gamma^{(n)}\Vert < \epsilon$. As a
measure for the deviation we use the entrywise norm $\Vert A\Vert=\sum_{ij}
\abs{a_{ij}}$. In principle it is possible to solve the Bethe-Salpeter equation
by supermatrix inversion (which requires a decoupling of the singlet channel
into spin and charge channels). However, close to the instability one would be
faced with the intricate task of inverting ill-conditioned matrices.
When iterating the equation the instability is signaled by a decelerated
convergence. This is related to the fact that the leading eigenvalue of the
corresponding matrix tends to one, so that more and more diagrams in the ladder need to be taken into account in order to obtain an accurate result. If one is not interested in the susceptibility itself but only in the instability, this can be circumvented by locating the instability by finding the parameters for which the leading eigenvalue of the BSE-related eigenvalue problem becomes one.

Note that the convolution of the two Green functions in Eqns. (\ref{eq::bse1},\ref{eq::bse2}) involves a sum over $n=N^d$ k-vectors for each value of $\vc{q}$ and thus of the order of $\mathcal{O}(n^2)$ arithmetic operations. This becomes tedious already for relatively small lattice sizes. It is then possible to reduce the order down to $\mathcal{O}(n\log n)$ by performing a so-called fast convolution using fast Fourier transforms. The convolution is calculated for all vectors $\vc{q}$ simultaneously and stored in memory. It is also used for the calculation of the susceptibilities (see below). Efficient transforms are provided by standard packages\cite{fftw}.

\begin{figure}[t]
\begin{psfrags}
\psfrag{w1}{$\!\!\!\Omega,\vc{q}$}
\psfrag{w2}{$\Omega,\vc{q}$}
\psfrag{g1}{\large $\!\!\Gamma^{ph(0)}$}
\psfrag{s1}{$\sigma$}
\psfrag{s2}{$\sigma$}
\psfrag{s3}{$\sigma'$}
\psfrag{s4}{$\sigma'$}
\psfrag{x1}{ }
\psfrag{x2}{$\Omega,\vc{q}$}
\psfrag{x3}{ }
\psfrag{x4}{ }
\begin{center}
\includegraphics[scale=0.6,angle=0]{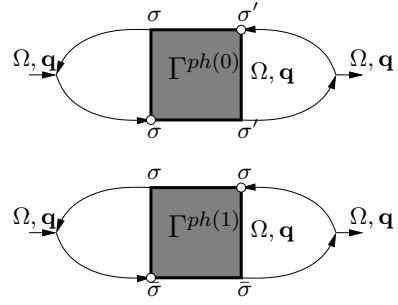}
\end{center}
\end{psfrags}
\begin{psfrags}
\psfrag{w1}{$\!\!\!\Omega,\vc{q}$}
\psfrag{w2}{$\Omega,\vc{q}$}
\psfrag{g1}{\large $\!\!\Gamma^{ph(1)}$}
\psfrag{s1}{$\sigma$}
\psfrag{s2}{$\bar{\sigma}$}
\psfrag{s3}{$\bar{\sigma}$}
\psfrag{s4}{$\sigma$}
\psfrag{x1}{ }
\psfrag{x2}{$\Omega,\vc{q}$}
\psfrag{x3}{ }
\psfrag{x4}{ }
\begin{center}
\includegraphics[scale=0.6,angle=0]{chi.eps}
\end{center}
\end{psfrags}
\caption{Diagrams for the susceptibilities in the two different particle-hole
channels}\label{fig::susc}
\end{figure}

Once we have found a converged solution for the two-particle vertex $\Gamma$,
we are able to calculate nonlocal susceptibilities.
For the paramagnetic case, we have
\begin{equation}
\langle S_z\, S_z\rangle(\Omega,\vc{q}) = \frac{1}{2}(\langle
n_{\uparrow}\, n_{\uparrow} \rangle - \langle
n_{\uparrow}\, n_{\downarrow}\rangle)(\Omega,\vc{q})\ ,
\label{eq::szsz}
\end{equation}
where $\langle
n_{\sigma}\, n_{\sigma'}\rangle(\Omega,\vc{q})=\chi_0^{\sigma\sigma'}(\Omega,\vc{q})
+\chi^{\sigma\sigma'}(\Omega,\vc{q})$. This is valid for lattice and dual fermions. For the dual fermions, the first term is given by the bubble
diagram (dual susceptibilities are marked by the tilde):
\begin{equation}
\tilde{\chi}_0^{\sigma\sigma'}(\Omega,\vc{q})=-\frac{1}{\beta N^d}\sum\limits_{\omega}
\sum\limits_{\vc{k}}G^{\text{d}\sigma}_{\omega}(\vc{k})G^{\text{d}\sigma'}_{\omega+\Omega}(\vc{k}+\vc{q})\ ,
\label{eq::bubble}
\end{equation}
while the nontrivial part of the susceptibility, cf. Fig. \ref{fig::susc}, is
given by
\begin{eqnarray}
\lefteqn{ \tilde{\chi}^{\sigma\sigma'}(\Omega,\vc{q}) =}\nonumber\\
&&\frac{1}{\beta^2 N^{2d}}\sum\limits_{\omega\omega'}\sum\limits_{\vc{k}\vc{k}'}
\Gamma^{ph0,\sigma\sigma'}_{\omega\omega'
\Omega}(\vc{q})G^{\text{d}\sigma}_{\omega}(\vc{k})G^{\text{d}\sigma}_{
\omega+\Omega}(\vc{k}+\vc{q})\times\nonumber\\
&&\hspace{10em} \times
G^{\text{d}\sigma'}_{\omega'}(\vc{k}')G^{\text{d}\sigma'}_{\omega'+\Omega}(\vc{k}'+\vc{q})\ . \nonumber\\
\label{eq::susc1}
\end{eqnarray}

For the other channel, the nontrivial part of the susceptibilities $\langle
S^\pm S^\mp\rangle$ is given by $\langle c_{\sigma}^\dagger c_{\bar{\sigma}} c_{\bar{\sigma}}^\dagger c_{\sigma} \rangle(\Omega,\vc{q})
= \chi_0^{\sigma\bar{\sigma}} + \chi^{\sigma\bar{\sigma}}$. For the dual fermions,this is obtained via
\begin{eqnarray}
\lefteqn{ \tilde{\chi}^{\sigma\bar{\sigma}}(\Omega,\vc{q}) =}\nonumber\\
&&\frac{1}{\beta^2 N^{2d}}\sum\limits_{\omega\omega'}\sum\limits_{\vc{k}\vc{k}'}
\Gamma^{ph1,\sigma\bar{\sigma}}_{\omega\omega'\Omega}(\vc{q})
G^{\text{d}\bar{\sigma}}_{\omega}(\vc{k})G^{\text{d}\sigma}_{\omega+\Omega}(\vc{k}+\vc{q})\times\nonumber\\
&&\hspace{10em}\times
G^{\text{d}\sigma}_{\omega'}(\vc{k}')G^{\text{d}\bar{\sigma}}_{\omega'+\Omega}(\vc{k}' + \vc{q})\nonumber\ . \\
\label{eq::susc2}
\end{eqnarray}

In order to be able to compare with known results, we need to transform the dual susceptibility to the susceptibility for the original fermions. According to Eqn. (\ref{eq::realchi}), this can be achieved by first multiplying the vertex by four dual Green function legs to obtain the dual two-particle Green function and then multiplying by the functions $L$ and $R$ defined in Eqns. (\ref{eq::l},\ref{eq::r}).
In order to calculate the susceptibility $\langle S_z\, S_z\rangle(\Omega,\vc{q})$ this merely amounts to calculate the susceptibility according to Eqns. (\ref{eq::szsz}-\ref{eq::susc1}), but with the dual Green functions in Eqn. (\ref{eq::susc1}) replaced by the modified propagators
\begin{eqnarray}
\mathcal{G}^L&=&(\Delta_{\omega\sigma}-h_{\vc{k}\sigma})^{-1}g_{\omega\sigma}^{-1}\, G^{\text{d}}_{\omega\sigma} \nonumber\\
\mathcal{G}^R&=& G^{\text{d}}_{\omega\sigma}\, g_{\omega\sigma}^{-1}\,(\Delta_{\omega\sigma}-h_{\vc{k}\sigma})^{-1} \nonumber\\
\end{eqnarray}
(these are identical for a diagonal basis) and the Green functions in Eqn. (\ref{eq::bubble}) by the lattice Green function.

The DMFT susceptibility can be straightforwardly obtained since we already have the fully reducible vertex of the local impurity problem, $\gamma^{(4)}$, at our disposal. From this we can obtain the corresponding vertex in the spin channel as $\gamma^{(4)}_s=\gamma^{(4)}_{\uparrow\uparrow}-\gamma^{(4)}_{\uparrow\downarrow}$. The irreducible vertex of the impurity problem is then obtained by matrix inversion using the standard relation
\begin{equation}
(\gamma^{(4)}_{s\omega\omega^\prime \Omega})^{-1}=(\gamma^{\text{ir}}_{s\omega\omega^\prime \Omega})^{-1} - \chi^0_{\omega\Omega}\, \delta_{\omega\omega^\prime}\ ,
\end{equation}
where $\chi^0_{\omega\Omega} = -\frac{1}{\beta}G^{\text{D,loc}}_\omega G^{\text{D,loc}}_{\omega+\Omega}$
and $G^{\text{D,loc}}_\omega=\sum_\vc{k} G^{\text{D}}_\omega(\vc{k})$ is the local DMFT Green function.
The vertex is obtained by iterating the BSE
\begin{eqnarray}
\Gamma^s_{\omega\omega' \Omega} (\vc{q}) =
\gamma^{\text{ir}}_{s\omega\omega' \Omega}&-&
\frac{1}{\beta N^d}\sum\limits_{\omega''}\sum\limits_{\vc{k}}\gamma^{\text{ir}}_{s\omega\omega'' \Omega} G^{\text{D}}_{\omega''}(\vc{k})
\times\nonumber \\
&&\times G^{\text{D}}_{\omega'' + \Omega}(\vc{k}+\vc{q})\, \Gamma^s_{\omega''\omega^\prime\Omega}(\vc{q})\nonumber\ ,\\
\label{eq::bseDMFT}
\end{eqnarray}
where $G^{\text{D}}$ is the DMFT lattice Green function, as before. The susceptibility itself is obtained using equations similar to Eqns. (\ref{eq::bubble},\ref{eq::susc2}) with the vertex and Green functions replaced by the appropriate DMFT quantities.

\subsection{Two-particle properties}

\begin{figure}[t]
\psfrag{a}{\tiny $(\pi,\pi)$}
\psfrag{b}{\tiny $(3/4\pi,\pi)$}
\psfrag{c}{\tiny $(3/4\pi,3/4\pi)$}
\psfrag{d}{\tiny $(0,0)$}
\psfrag{x}{\large $\tilde{\chi}^{zz}$}
\begin{center}
\vspace{2em}
\includegraphics[scale=0.8,angle=0]{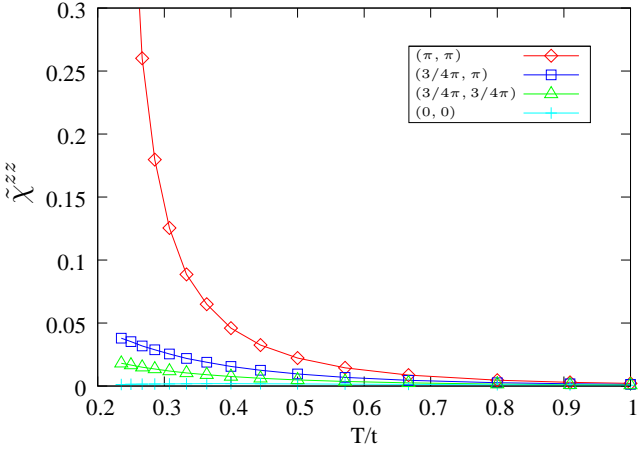}
\end{center}
\caption{(Color online) Nontrivial part of the dual susceptibility
$\tilde{\chi}^{zz}=\frac{1}{2}(\tilde{\chi}^{\uparrow\uparrow}-\tilde{\chi}^{
\uparrow\downarrow} )$ for different k-points as a function of temperature. The divergence at $\vc{q}=(\pi,\pi)$ indicates the antiferromagnetic instability.
}\label{fig::chidual}
\end{figure}

Now we turn to the two-particle properties. In order to illustrate the statement that two-particle excitations are the same for real and dual fermions, we first consider the dual susceptibility, Eqn. (\ref{eq::susc1}).
In Fig. \ref{fig::chidual}, we show the results for the nontrivial part $\tilde{\chi}^{zz}$ of the dual susceptibility as a function of temperature for $U=4$ and different k-points. This calculation was performed for an $8\times 8$ lattice. The susceptibility diverges at the wave vector $(\pi,\pi)$, indicating a transition to the antiferromagnetic ordered state, while at the other k-points it does not show any indication for a divergence. The susceptibility of the lattice fermions diverges at the same temperature, as expected. We would like to note that the dual bubble diagram, Eqn. (\ref{eq::bubble}) can be negative so that the dual susceptibility $\tilde{\chi}_0+\tilde{\chi}$ becomes negative at high temperatures. This is due to the fact that the spectrum of the dual Green function need not be positive-semidefinite, which follows from the fact that its local part exactly vanishes. Indeed, since the dual fermions represent the nonlocal part of the lattice fermions, they are no physical particles. However, this does not affect the lattice susceptibility as shown below and we did not encounter any non-analyticity problems in all our calculations.

\begin{figure}[t]
\begin{center}
\includegraphics[scale=0.182,angle=270]{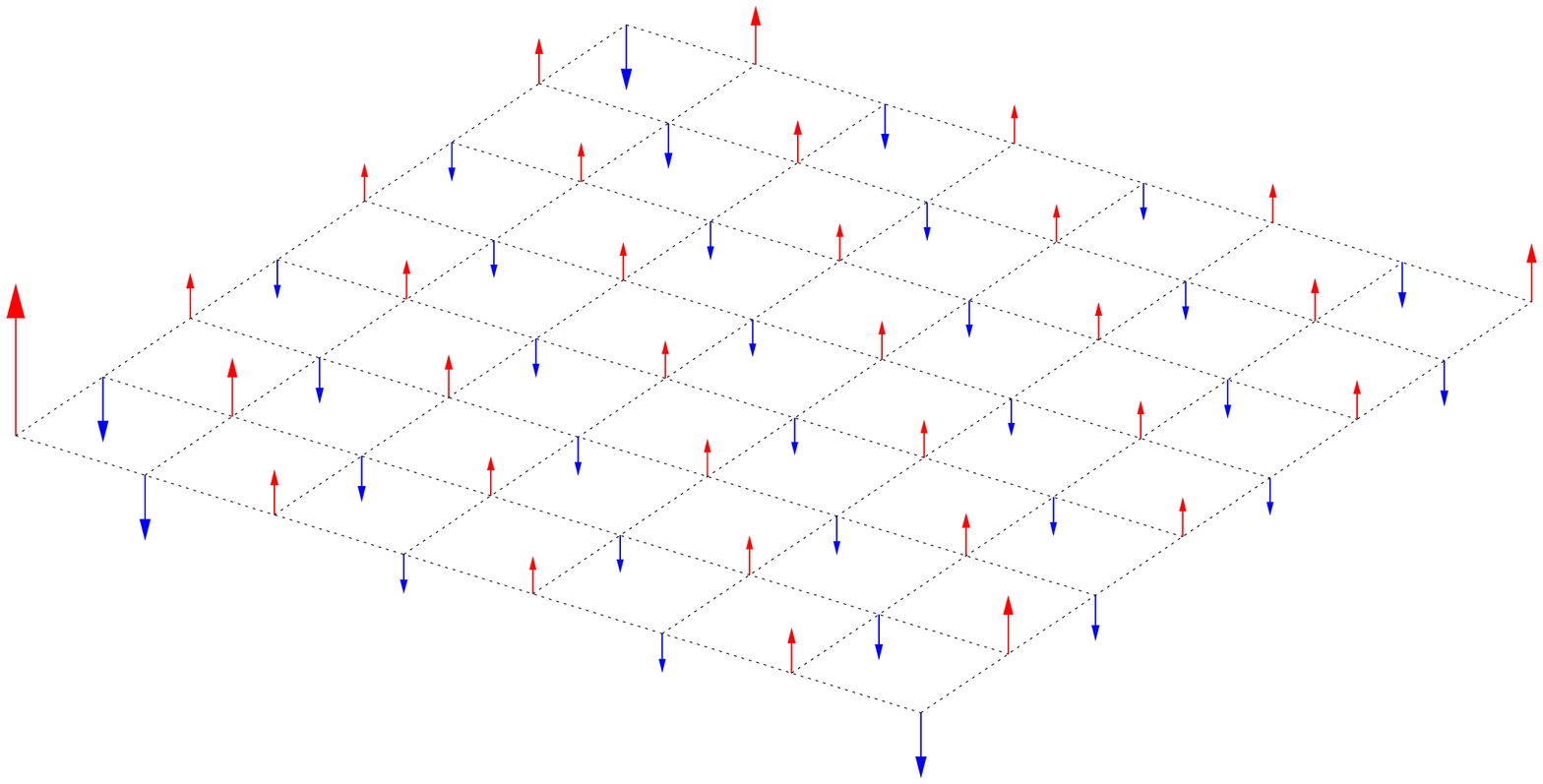}\hspace{-2em}\includegraphics[scale=0.182,angle=270]{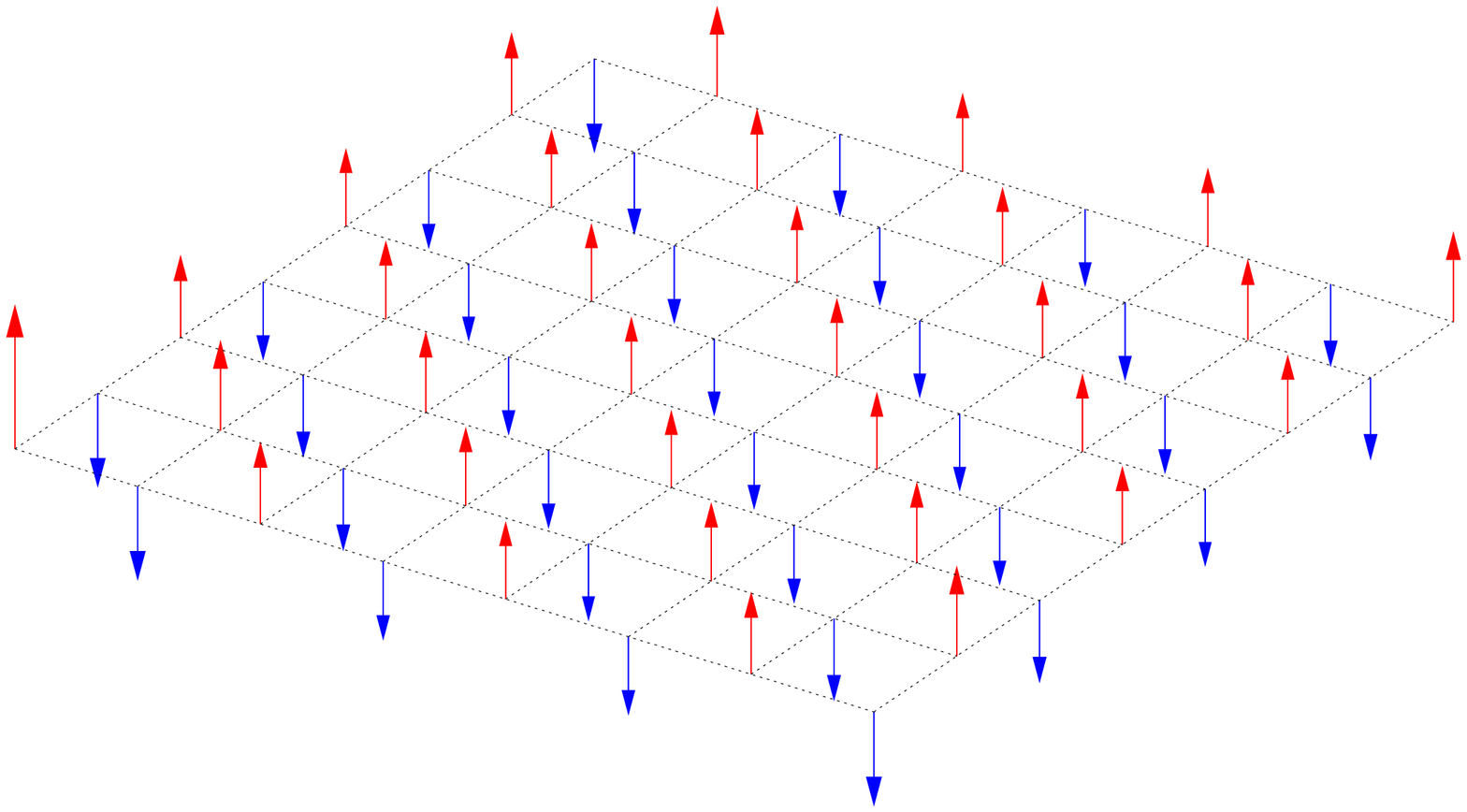}
\vspace{-2em}
\end{center}
\caption{(Color online) The antiferromagnetic correlations visualized in real space for $\beta=4.25$ and $\beta=4.5$. At temperatures close to the transition, the correlations become essentially independent of distance.
}\label{fig::chiR}
\end{figure}

In the paramagnetic case the relation $\frac{1}{2}(\langle S^+S^- \rangle +\langle S^-S^+ \rangle) = 2\langle S_z S_z \rangle$ holds. We find that this relation is very well fulfilled within our calculations. It follows from the invariance under rotations in spin space and it can be generally proven\cite{noz} that $\Gamma^-:=\Gamma^{ph0}_{\sigma\sigma\sigma\sigma} - \Gamma^{ph0}_{\sigma\sigma\bar{\sigma}\bar{\sigma}}$ (which describes the interactions of the particle-hole pair in the triplet state with spin projection $\sigma_z=0$) is equal to $\Gamma^{ph1}$ (which corresponds to the spin projections $\sigma_z=\pm 1$, depending on the sign of $\sigma$). This identity is readily shown to be preserved by the BSEs Eqns. (\ref{eq::bse1},\ref{eq::bse2}) given that  $\gamma_{\sigma\bar{\sigma}\bar{\sigma}\sigma} =
\gamma_{\sigma\sigma\sigma\sigma} - \gamma_{\sigma\sigma\bar{\sigma}\bar{\sigma}}$ holds, which is fulfilled in our calculations up to a small numerical error.

Using the definition of $\gamma^{(4)}$, Eqn. (\ref{eq::gamma4}), one finds
that this small numerical difference can be traced back to the error due to
Monte-Carlo (MC) averaging of the two-particle Green function: In each MC
measurement, one measures a quantity $\tilde{G}_{12}$ which corresponds to the
contribution of the particular configuration to Green's function (for details,
see Ref. \onlinecite{Rub05-1,Rub05-2,Rub05-3}). The Green function
$G_{12}=\langle\tilde{G}_{12}\rangle $ is obtained as the MC average of this
quantity. Similarly, using Wick's theorem, the two particle Green function
$\chi^{\sigma\sigma\sigma'\sigma'}_{1234}$ is obtained as
$\langle\tilde{G}^\sigma_{12}\tilde{G}^{\sigma'}_{34}\rangle -
\delta_{\sigma\sigma'}\langle\tilde{G}^\sigma_{14}\tilde{G}^\sigma_{32}\rangle
$. Thus one has $\chi^{\sigma\bar{\sigma}\bar{\sigma}\sigma}_{1234}=-\langle
\tilde{G}^{\sigma}_{14}\tilde{G}^{\bar{\sigma}}_{32}\rangle$ and
\begin{equation*}
\chi^{\sigma\sigma\sigma\sigma}_{1234}-\chi^{\sigma\sigma\bar{\sigma}\bar{
\sigma}}_{1234} =(\langle\tilde{G}^{\sigma}_{12}\tilde{G}^{\sigma}_{34}\rangle
-\langle\tilde{G}^{\sigma}_{14}\tilde{G}^{\sigma}_{32}\rangle) -
\langle\tilde{G}^{\sigma}_{12}\tilde{G}^{\bar{\sigma}}_{34}\rangle\ .
\end{equation*}
Since the quantities $\tilde{G}_{12}$ for different spins can differ even for
paramagnetic systems, the above quantities can only be equal within the MC
error.

In Fig. \ref{fig::chiR}, we show the real space dependence of the spin-spin susceptibility for the $8\times 8$ lattice. At lower temperatures, the correlations decay quickly as a function of distance, although the antiferromagnetic character is clearly visible. At lower temperatures close to the transition, the correlation length approaches the lattice size, and the susceptibility becomes essentially constant as a function of distance.

\begin{figure}[t]
\begin{center}
\includegraphics[scale=0.42,angle=0]{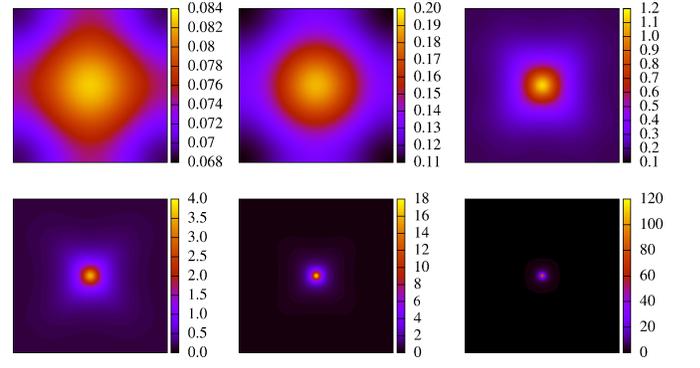}
\end{center}
\caption{(Color online) Momentum dependence of the static susceptibility (centered at the $M$-point) for $U=4$ and different temperatures calculated on a $256\times 256$ lattice. Top row from left to right: $\beta=0.5$, $\beta=1.0$ and $\beta=3.0$. Bottom row: $\beta=4.0$, $\beta=4.5$ and $\beta=4.65$.
}\label{fig::chiq}
\end{figure}

In Fig. \ref{fig::chiq} we illustrate the k-dependence of the susceptibility $\langle S_z S_z\rangle(\Omega=0,\vc{k})$. For low temperatures, the susceptibility appears to be delocalized in k-space, leading to the fast decaying correlations as a function of distance. However, even for the highest temperature the maximum is located at $\vc{k}=(\pi,\pi)$, exposing the tendency to antiferromagnetism. For low temperatures, the susceptibility becomes strongly peaked in the $(\pi,\pi)$ direction.

As far as the single-particle properties are concerned, our approach has proven to give physically correct results. However, since the calculation of the susceptibilities relies on an additional approximation, namely that we take the bare interaction of the dual fermions, $\gamma^{(4)}$, as an approximation for the irreducible vertex, we need to demonstrate that the approach still yields sensible results for the two-particle properties.

\begin{figure}[t]
\psfrag{x}{\large $T/t$}
\psfrag{y}{\large $1/\chi_{\text{AF}}$}
\begin{center}
\vspace{2em}
\includegraphics[scale=0.7,angle=0]{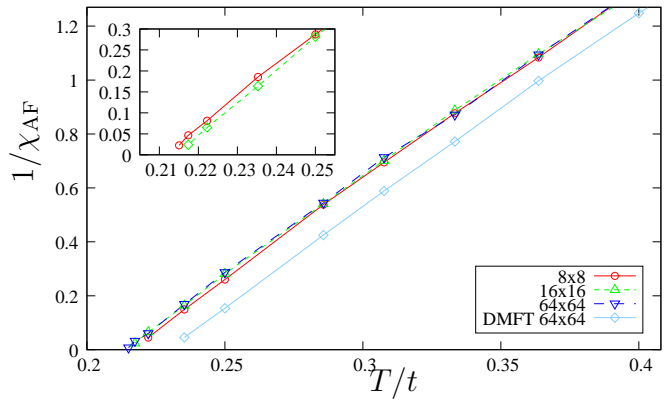}
\end{center}
\caption{(Color online) Inverse of the antiferromagnetic susceptibility $\chi_{\text{AF}}=\langle S_z\,S_z\rangle(\Omega=0,\vc{q}=(\pi,\pi))$ as a function of temperature as obtained from DMFT and within the dual fermion approach for $U=4$ and different lattice sizes. Here e.g. the label $64\times 64$ indicates the number of k-points used in the Brillouin zone integration. The inset compares the fully self-consistent DF result (dashed line) with the one obtained from the first iteration of the outer loop (solid line) shown in Fig. \ref{fig::procedure}.
}\label{fig::chiU4}
\end{figure}

\begin{figure}[t]
\psfrag{x}{\large $T/t$}
\psfrag{y}{\large $1/\chi_{\text{AF}}$}
\begin{center}
\vspace{2em}
\includegraphics[scale=0.7,angle=0]{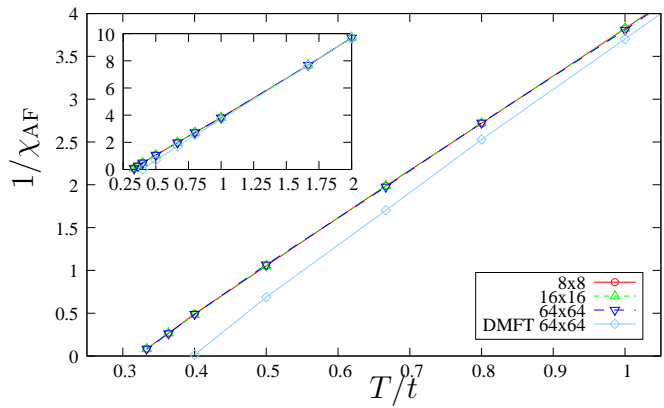}
\end{center}
\caption{(Color online) The antiferromagnetic susceptibility for $U=8$. The inset shows the same on a larger scale. For high temperatures, the dual fermion result converges to the DMFT result.
}\label{fig::chiU8}
\end{figure}

A stringent test for this approximation is whether the results are still an improvement compared to the DMFT. In Figs. \ref{fig::chiU4},\ref{fig::chiU8} we therefore plot the inverse antiferromagnetic susceptibility $\chi_{\text{AF}}=\langle S_z S_z\rangle(\Omega=0,\vc{k}=(\pi,\pi))$ as a function of temperature. The critical temperature is given by the point to which the corresponding curve extrapolates to zero. Note that the exact solution is expected to have a N\'eel temperature $T_N=0$. One might expect that since the DF and DMFT results are rather similar for $U=4$, this also applies for the susceptibility. Still, we find that the DMFT result is systematically below the DF result, and the DF critical temperature is correspondingly lower than in the DMFT. This can be interpreted as an effect of the non-local spin fluctuations which are not accounted for in the DMFT and which suppress the critical temperature.
In order to understand the scaling behavior one should recognize that unlike the DCA, which essentially is an expansion in $1/N_c$ ($N_c$ is the clustersize) and hence becomes formally exact in the limit of infinite clusters\cite{hettler00,maier05}, our approach does not become exact in the limit of infinite lattice sizes. Thus we cannot reproduce the scaling behavior as observed in DCA calculations\cite{jarrell01}. This is due to the termination of the perturbation series. On the other hand, our approach becomes formally exact if all diagrams are considered in the limit of infinite lattice sizes. Since we sum a perturbation series for auxiliary fermions, an \emph{a priori} statement which diagrams include what kind of physics in terms of the lattice fermions is not possible. However, we find that the approach is essentially converged for lattices larger than $16\times 16$, setting a scale for the range of the fluctuations included. We expect this scaling to be more significant if either the number of diagrams considered is increased, or the starting point is improved, i.e. the dual fermion calculation is performed on top of an e.g. $2\times 2$  cluster-DMFT calculation.

The inset of Fig. \ref{fig::chiU4} compares the antiferromagnetic susceptibility obtained from a fully self-consistent DF calculation (dashed line, with diamonds) to that obtained from the first iteration of the outer loop, i.e. just one inner loop as shown in Fig. \ref{fig::procedure} has been performed. The hybridization is thus the one obtained from DMFT. One sees that the fully converged DF result has a slightly larger critical temperature than the intermediate result. We have observed this tendency in all our calculations. We attribute this to the renormalization of the hybridization as compared to DMFT, which pronounces the effect of the local singlet formation and thus favors antiferromagnetism.

In Fig. \ref{fig::chiU8} we show the temperature scaling of the inverse susceptibility for $U=8$. Here we find that almost no scaling with the lattice size is visible. This can be attributed to the fact that the physics of singlet formation prevails in this parameter regime and this is what is primarily mediated by the first diagram in the perturbation expansion. This is consistent with the fact that the transition temperature is higher compared to $U=4$. We thus expect a considerable reduction of the transition temperature when the local singlet-formation is treated explicitly within the cluster-extension of our approach. However, the fluctuations are still taken into account and lead to a suppression of the critical temperature which is somewhat more pronounced than for $U=4$.

\section{Conclusions}

To conclude, we have proposed a scheme to calculate the two-particle Green
function within the dual fermion framework. This enables us to study
two-particle interactions in strongly correlated systems where non-local correlations cannot be neglected. We formulated the Bethe-Salpeter equations for the full vertex in the particle-particle and particle-hole channels and proposed an approximation for practical calculations.  A possible extension of this scheme has been
proposed which allows the application to superconductivity while taking also the
antiferromagnetic fluctuations into account. 
We have established an exact relation between the two-particle Green functions in dual and original variables, which ensures that two-particle excitations of real and dual variables are identical. The identity was used to transform the dual susceptibility to the one of the original fermions. Within our proposed approximation we have applied the scheme to the 2D Hubbard model at half filling in the paramagnetic phase. We find strong modifications of the single-particle properties in our dual fermion calculations compared to the DMFT. This is encouraging given that our calculations were performed within the single-site formalism. Concerning the two-particle properties, we also found an improvement compared to the DMFT. The critical temperature is suppressed due to the incorporation of the non-local spin fluctuations, thus showing that this approximation goes well beyond the conventional DMFT scheme for calculating the two-particle Green function. We expect a further improvement of the results by the cluster formulation of our approach, which explicitly takes the local singlet formation into account.

\begin{acknowledgments}
This work was supported by NWO project 047.016.005 and FOM (The
Netherlands), DFG Grant No. SFB 668-A3 (Germany), the Leading
scientific schools program and the ``Dynasty'' foundation
(Russia) and partly by the National Science Foundation under
Grant No. PHY05-51164. We also would like to thank H. Monien and F. Anders for
fruitful discussions.
\end{acknowledgments}

\bibliography{cond_matv2}

\end{document}